\documentclass[12pt,a4paper]{article}
\usepackage{times}
\usepackage[utf8]{inputenc}
\usepackage{graphicx}
\usepackage{url}
\usepackage[pdftex]{color}
\usepackage{epstopdf}
\usepackage{cite}

\renewcommand{\vec}[1]{{\mathbf #1}}
\newcommand{\sgn}{{\rm sign}}

\newcommand{\kparallel}{k_\parallel}
\newcommand{\veckparallel}{\vec{k}_\parallel}
\begin{document} 

\title{\bf
Cherenkov friction on a neutral particle moving parallel to a dielectric%
}
\author{%
Gregor Pieplow and
Carsten Henkel%
\\
\normalsize
\textit{Institute of Physics and Astronomy, Universit\"at Potsdam,
Germany}%
}
\date{\normalsize 18 Feb 2014}
\maketitle

\begin{abstract}
\noindent
Based on a fully relativistic framework and the assumption of
local equilibrium, we describe a simple mechanism of quantum friction for a particle moving parallel to a dielectric. The Cherenkov effect explains how the bare ground state becomes globally unstable and how fluctuations of the electromagnetic field and the particle's dipole are converted into pairs of excitations. Modelling the particle as a silver nano-sphere, we investigate the spectrum of the force and its velocity dependence. We find that the damping of the plasmon resonance in the silver particle has a relatively strong impact near the Cherenkov threshold velocity. We also present an expansion of the friction force near the threshold velocity for both damped and undamped particles.
\end{abstract}

\section*{Introduction}

\noindent
The conversion of mechanical energy into heat is  
referred to as friction in most cases. Numerous mechanisms can be identified 
that cause friction, but it is still a challenge to infer macroscopic 
observations from microscopic phenomena. 
So far only very simple scenarios permit a detailed analysis 
of the fundamental aspects of friction. 
A prominent example is the theory of the quantized electromagnetic 
field applied to the case of two parallel moving plates 
separated by a small vacuum gap \cite{teodorovich1978,levitov1989van,polevoi1990tangential,pendry1997shearing,philbin2009no};
see Refs.\cite{kardar99friction,davies2005quantum,volokitin2007near,maghrebi2013bquantum} for reviews. 
Friction arises due to the spontaneous creation of particle pairs
that propagate away into the plates or are dissipated there.
A similar treatment can be applied to a body moving above a flat surface
at constant speed \cite{schaich81dynamic,dedkov1999electromagnetic,scheel2009,barton2010van2}. 
Taking advantage of Lorentz invariance, one achieves treatments consistent with special relativity \cite{kyasov2002relativistic,dedkov2009fluctuation},
as required for the archetypal situation that high-energy charges are stopped 
in a medium.
In a recent paper, we described such a formalism for a neutral,
polarizable 
particle moving parallel to a flat interface \cite{pieplow2013fully}. 
At a typical distance of at least a few nanometers (larger than the
atomic scale), 
the interaction depends on a few macroscopic parameters (refractive index, conductivity, surface impedance \ldots). 
In the present paper we discuss a special configuration of this 
setting with a dielectric medium below the surface, and with both particle and 
medium at zero temperature. 
These conditions make the friction a pure quantum-mechanical drag 
and closely relates it to the realm of Casimir phenomena \cite{dalvit2011fluctuations}. 
A friction force appears when the speed of the particle 
relative to the surface exceeds the velocity of light in the medium
($c/n$): 
this drag can thus be attributed to the Cherenkov effect. The situation
is somewhat unusual because neither the surface nor the particle have to 
be dissipative. All that is required are spectral mode densities for
the medium field and the particle. 
For a moving charge the Cherenkov drag is well known and is described 
easily with 
classical electromagnetic theory \cite{bolotovskii1957theory}.
A neutral body requires a more refined treatment, as quantum fluctuations 
have to be treated accordingly.
As in previous work \cite{polevoi1990tangential,rytov1989elements}, 
we use the fluctuation-dissipation theorem to do just that.
Because of the growing interest in this field and some controversy 
surrounding it (see Ref.\cite{barton2010van2} for a review),
the simple situation studied here might provide another test bed 
to compare current results and ideas in detail.

In this paper, we analyze in detail a spectral representation
of the friction force that must be applied to move a small particle
parallel to a flat dielectric surface. While this setup has obvious
applications for micro- and nano-machines, our focus is on illustrating
the underlying mechanisms. The basic physics is very similar to the 
seminal explanation of the Cherenkov effect \cite{cherenkov1937visible}
by Tamm and Frank \cite{tamm1937coherent}: for a certain sector of
field modes, the Doppler shift flips the sign of the mode frequency
(anomalous Doppler effect).
This leads to scattering relations (S-matrix) in the form of a Bogoliubov
transformation \cite{davies2005quantum,maghrebi2013bquantum}:
incident waves get amplified, 
and pairs of elementary excitations (photon-polaritons) can be created
out of the quantum fluctuations in the field and in the particle's 
dipole moment. The
frictional force arises from the power carried away by these
excitations as they are absorbed or as they propagate into the bulk of 
the body. The recent paper by Barton \cite{barton2010van2} provides a
particularly transparent calculation of these processes in a simplified
setting (only surface plasmon modes are considered). 
The starting point we use here is based on the fluctuation electrodynamics
developed by Rytov and co-workers \cite{rytov1989elements}: the basic
assumption is that both the solid surface and the moving particle are
in \emph{local thermodynamic equilibrium}. This is
a good approximation for a mesoscopic particle made from thousands
of atoms, at least over time scales where its temperature can be
considered constant (large heat capacity). The approximation is much 
more questionable for microscopic particles like atoms or molecules 
because these may settle into a non-thermal state due to spontaneous 
excitation.

We structure our analysis in the following way:
some results of previous work are summoned to provide the basis for the 
Cherenkov effect. 
The quantum (Cherenkov) friction is then calculated and its
physical properties are discussed.
We link the friction force to an absorbed power that has to be
provided to move the particle at constant speed. A relativistic
argument put forward by Polevoi \cite{polevoi1990tangential} 
attributes this power to an increase in mass-energy.
After the analytics, we numerically investigate a silver nano-particle 
moving at relativistic speed above a dielectric surface. We quantify 
the magnitude of the friction, and provide a geometric picture of most 
of the features that determine the friction spectrum. 
We then present an expansion of frequency spectrum of the force and 
of the force itself near the threshold in $(v-c/n)$. This further illustrates the relative importance of the resonance and 
the low, off-resonant frequencies in the particle polarizability. 
We find a remarkable agreement with the numerical results close to the threshold. 
The main result is that the Cherenkov friction is linked to composite
modes at the vacuum-dielectric interface \cite{carnaglia1971quantization,eberlein2006quantum} which couple to the particle via
their evanescent vacuum tail; their plane-wave component
in the medium can be seen as carrying away the dissipated power.

%
\section{The formalism}
%
%
\subsection{Friction force}

In an earlier paper \cite{pieplow2013fully} we presented a covariant 
approach to the force on a particle that moves with arbitrary speed 
parallel to a flat surface that responds linearly to electromagnetic
waves.
We recovered the results of Refs.\cite{mkrtchian1995interaction,kyasov2002relativistic}.
The formalism allows for different temperatures of particle and surface,
assuming a state of local equilibrium. The relative motion leads to
Doppler shifts that are handled by Lorentz transforming an incident
field into the frame co-moving with the particle or the surface.
The Doppler-shifted frequency distribution of the equilibrium distributions 
are responsible for a non-equilibrium force that persists even when both
temperatures $T \to 0$. 

Let us fix coordinates such that the $x$-axis points along the motion
of the particle (velocity $\vec{v}$), while the half-space $z \le 0$ coincides 
with the medium. 
According to Refs.\cite{kyasov2002relativistic,pieplow2013fully}, 
the force component $F_x$ acting on the particle 
(at distance $z$ from the surface) is
\begin{eqnarray}
F_x &=& \frac{\hbar}{2 \gamma}
			\int\frac{{\rm d}\omega}{2 \pi}
			\frac{{\rm d}^2 \kparallel}{(2\pi)^2}
			[\sgn(\omega)-
			 \sgn(\omega-v k_x)]\times \nonumber
			\\
	&& \qquad k_x\,
			\mathop{\rm Im}\,\alpha[ \gamma ( \omega-v k_x ) ]
			\sum_{\sigma=s,p}\phi_{\sigma}(\omega,\veckparallel)
			\mathop{\rm Im}\left(\frac{r_\sigma {\rm e}^{-2 \kappa z}}{\kappa}\right)
	~.
\label{eq:generalForce}
\end{eqnarray}
The frequency $\omega$ and parallel wave numbers $\veckparallel 
= (k_x, k_y)$ are measured in the rest frame of the medium; 
the integral boundaries are $(-\infty,\infty)$. The
difference of sign functions arises from the thermal factors
$\coth[ \hbar \omega / (2 k_{\rm B} T ) ]$ in the zero-temperature
limit, evaluated in the respective rest frames of medium and
particle ($\omega' = \gamma (\omega - v k_x)$).
The particle polarizability is $\alpha$, $\gamma$ is the Lorentz 
factor, and for the weight functions 
$\phi_\sigma$ we have (setting $c = 1$)
\begin{eqnarray}
\phi_s(\omega,\veckparallel)
&=& 
	\omega^{\prime 2}  
	+ 
	2\gamma^2(\vec{v}\times\veckparallel)^2
	\left(1 - \frac{\omega^2}{\kparallel^2} \right)
	~,
	\\
\phi_p(\omega,\veckparallel)	
&=& 
	\omega^{\prime 2} 
	+
	2\gamma^2(\kparallel^2 - 
	(\vec{v}\cdot\veckparallel)^2)
	\left(1-\frac{\omega^2}{\kparallel^2}\right)
	~.
\end{eqnarray}
The reflection coefficients for $p$- and $s$-polarized light are 
\begin{equation}
r_{\rm s} = \frac{{\rm i}\kappa - \kappa_n}{{\rm i}\kappa + \kappa_n}
\quad,\quad 
r_{\rm p} = \frac{{\rm i} n^2\kappa - \kappa_n}{{\rm i} n^2\kappa + \kappa_n}
~,
	\label{eq:reflection-coefficients}
\end{equation}
where $\kappa = \sqrt{\kparallel^2 - (\omega+i0)^2}$ and  
$\kappa_n = \sqrt{n^2(\omega+i0)^2 - \kparallel^2}$.
Here, $n$ is the refractive index of the medium.

\begin{figure}[h!]%
\begin{tabular}{c@{\hspace*{1em}}c}
	 \includegraphics[width=7cm,angle=0,origin=c]{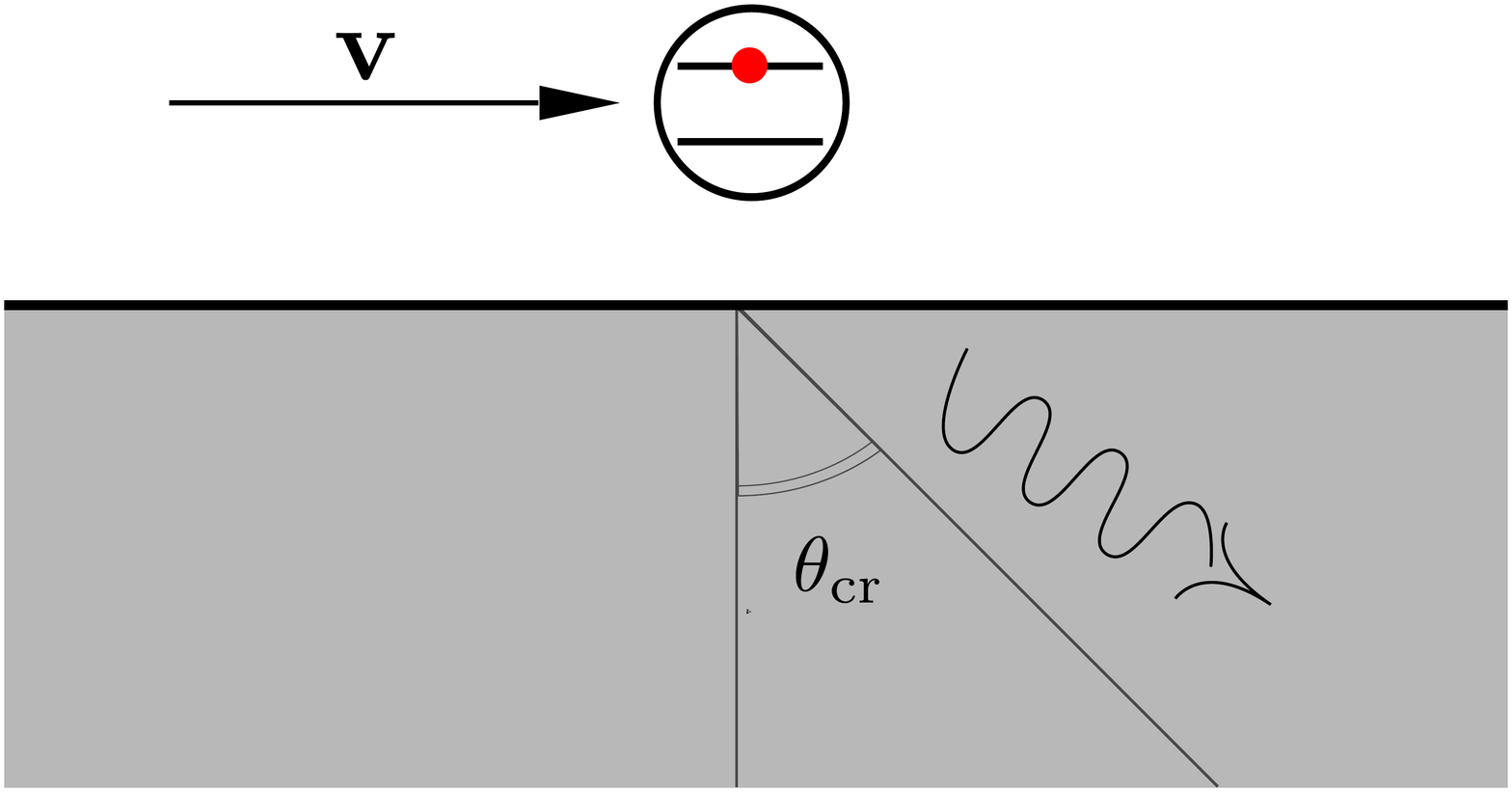}
&
	\includegraphics[width=7cm]{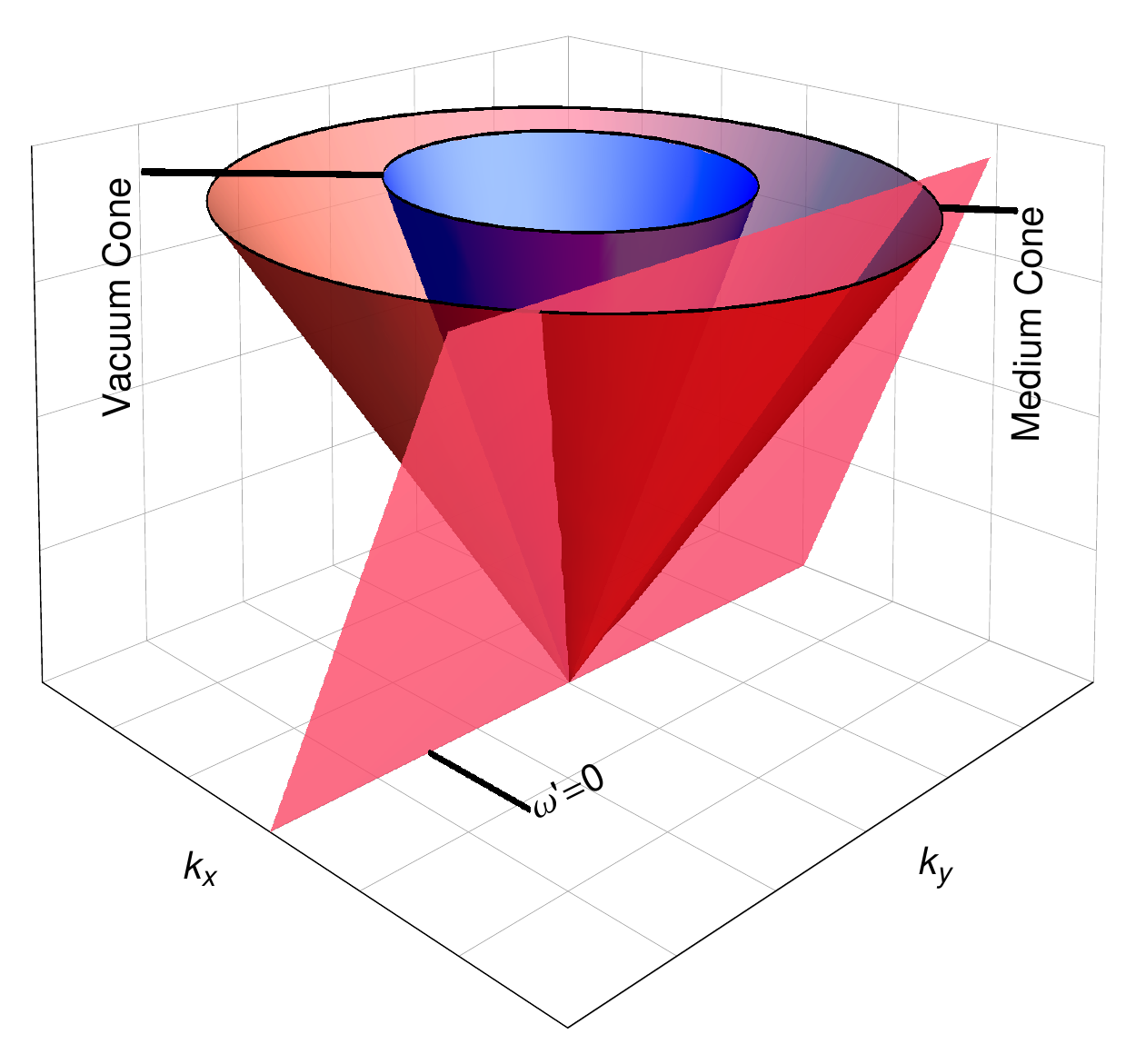}
\\
\multicolumn{2}{l}{%
\begin{minipage}[t]{\textwidth}
\refstepcounter{figure}%
\footnotesize{\bf Figure~\thefigure.}
Sketch of the system and visualization of the relevant photon
modes.
\label{fig:conesAndSlices}%
\end{minipage}}
\\
\begin{minipage}[t]{0.5\textwidth}\footnotesize
(\emph{left})
Kinematics of Cherenkov (quantum) friction: 
the moving particle is spontaneously excited 
and a photon is emitted into the medium beyond the critical angle.
\end{minipage}
&
\begin{minipage}[t]{0.5\textwidth}\footnotesize
(\emph{right})
Cherenkov friction arises from a domain in the $\omega,\veckparallel$-space
that is enclosed by the projected light cone in the medium 
$\omega = \kparallel / n$ (dark red) and the plane 
$\omega = v k_x$ (pink).
Below this plane, the Doppler shift is anomalous and in the frame
moving with the particle $\omega' < 0$.
All points below the vacuum light cone (blue) correspond to evanescent
waves bound to the medium surface.
We take $n = 2$ and $v = 0.8 \, c > c/n$.
	\label{fig:coneIntDomain}
\end{minipage}
\end{tabular}
\end{figure}

Using symmetries and other properties we can further simplify the integral in Eq.(\ref{eq:generalForce}). The integrand is even under the transformations 
$(\omega,k_x) \mapsto (-\omega,-k_x)$ 
and $k_y \mapsto -k_y$
so that it is sufficient to integrate over 
the domain $\omega > 0$, $k_y>0$. 
The difference of the $\sgn$-functions reduces to a factor of two for 
$0 < \omega < v k_x$. This wedge-shaped domain in the $k_x,\omega$-plane 
is below the projected light cone $\omega = \kparallel$,
so that only fields that are evanescent at the particle's location
contribute to the force [Fig.\ref{fig:conesAndSlices}(\emph{left})].
We conclude that the factor ${\rm e}^{-2\kappa z} / \kappa$ is real-valued.

Another crucial insight is contained in the reflection 
coefficients~(\ref{eq:reflection-coefficients}):
their imaginary part is nonzero only in the annulus
$\omega < \kparallel < n \omega$
[see Fig.\ref{fig:impact-resonance}(\emph{right}) below].
With the condition derived from the $\sgn$ functions we get 
$\omega < v \kparallel \cos \phi < v n \omega \cos \phi$,
so that the condition for Cherenkov radiation follows
\begin{equation}
1 < v n \cos \phi
	\label{eq:Cherenkov-condition}
\end{equation}
where $\phi$ is the angle between $\vec{v}$ and $\veckparallel$.
The expression~(\ref{eq:generalForce}) for the force thus becomes:
\begin{eqnarray}
F_x &=& \frac{4 \hbar}{\gamma (2 \pi)^3 }
	\int^\infty_0 {\rm d}\omega 
	\int^{n \omega}_{\omega/v}{\rm d}k_x
	\int^{\sqrt{n^2 \omega^2 - k_x^2}}_{0}{\rm d}k_y \nonumber
	\\
	&& \quad
	k_x\,
	\mathop{\rm Im}\,\alpha[ \gamma ( \omega - v k_x ) ]
	\sum_{\sigma=s,p}\phi_{\sigma}(\omega,\veckparallel)
	\mathop{\rm Im}\left( r_\sigma \right) \frac{{\rm e}^{-2 \kappa z}}{\kappa}
	~.
\label{eq:generalForceKx}
\end{eqnarray}

\subsection{Photon emission and anomalous Doppler shift}

The manipulations performed so far have a clear physical meaning
within the theory of the Cherenkov effect \cite{cherenkov1937visible,tamm1937coherent,bolotovskii1957theory}
which is well understood.
A kinematic explanation of the friction above the Cherenkov threshold 
can be given following the equations outlined in Ref.\cite{frolov1986excitation}. 
We start with the conservation of 4-momentum
\begin{equation}
p^\mu_1 = \hbar k^\mu + p^\mu_2
~. \label{eq:momentumConservation}
\end{equation}
The momenta $p^\nu_a$ describe the particle before and after 
the emission of a photon with momentum $\hbar k^\nu$, where 
$a = 1, 2$ labels the internal states (energy levels $\epsilon_{1,2}$).
Although Eq.(\ref{eq:momentumConservation}) and Ref.\cite{frolov1986excitation}
deal with a particle moving through a medium, the physics is the same
for the motion parallel to the dielectric medium.
We have for the particle and the photon (recall that $c = 1$)
\begin{eqnarray}
p^\mu_a &=& (E_a\,,\gamma m_a\vec{v})
~,\quad 
m_a = M + \epsilon_a
	~,
\\ 
E_a &=& \sqrt{m_a^2  + \gamma^2 m_a^2 \vec{v}^2} = \gamma m_a
	~,
\\
k^\mu &=& (\omega, \,\vec{k})
	~,\quad k = n\omega 
	~. 
\label{eq:dielectricDispersion}
\end{eqnarray}
The Greek indices run from $0$ to $3$, and 
toggling between co- and contravariant indices is done with the 
metric $g_{\mu \nu} = \mathop{\rm diag}(1,-1,-1,-1)$.
It is understood that $k = \sqrt{\vec{k}^2}$.
The masses $m_a$ are associated with the particle's energy levels.
The photon is supposed to be emitted into the medium, hence 
the dispersion relation in eq.(\ref{eq:dielectricDispersion}). 
Because the particle is pushed by an ``invisible hand", the velocity
$\vec{v}$ does not change during the emission. This is equivalent to 
neglecting the recoil~\cite{frolov1986excitation} of the particle. 
Squaring eq.(\ref{eq:momentumConservation}) leads to 
\begin{equation}
(\epsilon_1 - \epsilon_2)(2 M + \epsilon_1 + \epsilon_2) 
= 2 E_1\hbar \omega (1 - v n \cos \phi)
~.
\label{eq:energyImpulsFull}
\end{equation}
with the same notation as in Eq.(\ref{eq:Cherenkov-condition}) above.
We can reasonably make the approximation
$\epsilon_{1,2} \ll M$ so that we recover 
\begin{equation}
\hbar \omega = - \frac{ \epsilon_2 - \epsilon_1
	}{ \gamma(1 - v n \cos \phi) }
	~.
	\label{eq:emitted-photon-energy}
\end{equation}
If the particle is faster than 
the speed of light inside the medium, $1/n$, 
the denominator is negative [Cherenkov 
condition~(\ref{eq:Cherenkov-condition})].
This is an illustration of the so-called anomalous Doppler effect 
where the photon frequency, as seen from
the moving particle, $\omega' = \gamma(\omega - v k_x)$, is negative. 
The authors of Ref.\cite{frolov1986excitation} point out that
this allows for the 
\emph{excitation of the particle to a higher energy level},
$\epsilon_2 > \epsilon_1$,
\emph{while emitting a photon into the medium},
inside the Cherenkov cone [see 
Fig.\ref{fig:conesAndSlices}(\emph{left})].
The power lost into the emission must be supplied by the force 
that keeps the particle on its track. 
In other words, considering quantum electrodynamics at 
a dielectric interface coupled to a polarizable particle moving faster
than the Cherenkov threshold, it turns out that this
is an example of an unstable field theory \cite{davies2005quantum,silveirinha2013aquantization},
similar to electron-positron production in strong electric fields
and Hawking radiation in a strong gravitational field.

\subsection{Heating and frictional power}

This simple kinematic analysis corresponds neatly to the integration 
domain in eqs.(\ref{eq:generalForce}, \ref{eq:generalForceKx}). 
Note in particular that the 
particle's response function is evaluated at the Doppler-shifted frequency 
and yields
$\mathop{\rm Im} \alpha(\gamma(\omega - v k_x)) < 0$ 
in the domain. 
This is a clear indicator that the anomalous Doppler effect 
in combination with the photon emission of photons into the Cherenkov cone 
indeed slows down the particle. 
Another quantity of interest is the rate of mass change in the
particle's co-moving frame. This is given by
$\dot{m} = u^\mu F_\mu$ where $u_\mu$ is the particle's 4-velocity.
The full 4-vector of force $F_\mu$ 
can be found in \cite{pieplow2013fully},
and for our particle moving in the $x$-direction, we find
\begin{eqnarray}
\dot{m} &=& \gamma \left(F_0 - v F_x\right) 
\\
\left( {F_0 \atop v F_x} \right)
        &=& 
        \int_0^\infty \!{\rm d} \omega\, 
		  \int^{n \omega}_{\omega/v} \!{\rm d}k_x\,
		  \int^{\sqrt{n^2 \omega^2 - k_x^2}}_{0}\!{\rm d}k_y\,
\left({ - \hbar \omega \atop
	- \hbar v k_x }\right)
	\Gamma(\omega,\veckparallel)
~,
	\label{eq:increase-mass-2}
\end{eqnarray}
where the positive quantity
\begin{eqnarray}
\Gamma(\omega,\veckparallel)& = \frac{4 }{\gamma (2 \pi)^3}
	{\rm Im}\,\alpha[\gamma ( v k_v - \omega )]
	\sum_{\sigma=s,p}\phi_{\sigma}(\omega,\veckparallel)
	\mathop{\rm Im} (r_\sigma ) \frac{{\rm e}^{-2 \kappa z}}{\kappa}
\end{eqnarray}
can be identified as a spectrally resolved photon emission rate. 
(We exploited the fact that $\mathop{\rm Im}\alpha( \omega' )$ is an odd
function.)
Note that the proper mass increases, $\dot{m} > 0$, because per
emission event, a positive energy $- \hbar\omega' 
= \hbar\gamma( v k_x - \omega )$ 
is dumped into the particle's internal mass-energy,
as discussed in the previous section. Indeed, we shall see, through a 
simple oscillator model for the polarizability, that the
frequency $\omega'$ in the co-moving frame
is essentially fixed by the particle's resonance.

To summarize this section, let us re-write the power balance
as a sum of two positive terms:
\begin{equation}
- v F_x = - F_0 + \frac{ {\rm d} m }{ \gamma {\rm d}\tau }
\label{eq:emitted-power-and-absorbed-heat}
\end{equation}
On the left-hand side, we see the frictional power spent to 
maintain the constant speed of the particle. The first term
on the right-hand side gives the power of photon emission
(energy $\hbar\omega$ at rate $\Gamma( \omega, \veckparallel )$,
see Eq.(\ref{eq:increase-mass-2})),
while the second gives the power absorbed in the particle.
(The factor $1/\gamma$ gives the relativistic time dilation
between the particle's proper time $\tau$ and the laboratory
time $t$.)

\section{Case study: relativistic nanoparticle}

\subsection{Numerical investigations}
\label{s:plots}

To illustrate further the physical features of the Cherenkov friction
force, we provide some numerical estimates for a metallic nano-particle.
We chose a silver nano-sphere with radius $a = 3\,{\rm nm}$
that moves at a distance $z = 10\,{\rm nm}$ 
above a dielectric 
medium with refractive index $n = 2$. For simplicity, 
frequency dispersion is neglected in the 
medium \cite{eberlein2006quantum}.
For the particle, we adopt a Drude model with parameters for 
silver: plasma frequency
$\hbar\omega_{\rm pl} = 9.01 {\rm eV}$ and damping rate
$\hbar / \tau = 16 \,{\rm meV}$
(not to be confused with the proper time coordinate $\tau$ above).
For such a small particle, the first term of the Mie series will
suffice so that its response is given by the electric dipole
polarizability (for SI units, multiply with the Coulomb constant
$\varepsilon_0$)
\begin{equation}
\alpha(\omega) = 4\pi a^3 
\frac{\varepsilon(\omega) - 1}{\varepsilon(\omega) + 2}
=
4\pi a^3 
\frac{ \Omega^2 }{\Omega^2 - \omega^2 - {\rm i} \omega / \tau}
	\label{eq:particle-polarizability}
\end{equation}
where $\varepsilon( \omega )$ is the metal permittivity.
The resonance at $\Omega = \omega_{\rm pl} /\sqrt{3}$
corresponds to a plasmon mode localized on the particle.
The calculations simplify considerably in the no-damping limit
$\tau \to \infty$ which gives
\begin{equation}
\lim_{\tau\rightarrow\infty} \mathop{\rm Im} 
\alpha(\omega)
= 
2\pi^2 a^3\omega\,
[\delta(\omega - \Omega) + \delta(\omega + \Omega)]
~.
	\label{eq:imagAlphaTauInf}
\end{equation}
We have checked that at this distance and for velocities above
the Cherenkov threshold, both polarizations contribute roughly the same amount to the force. This is at variance with the more
familiar regime of short (non-retarded) distances and slow
(non-relativistic) atoms where the p-polarization dominates
and an electrostatic calculation suffices.
\begin{figure}[h!]%
\begin{tabular}{c@{\hspace*{1em}}c}
	\includegraphics[width=7cm]{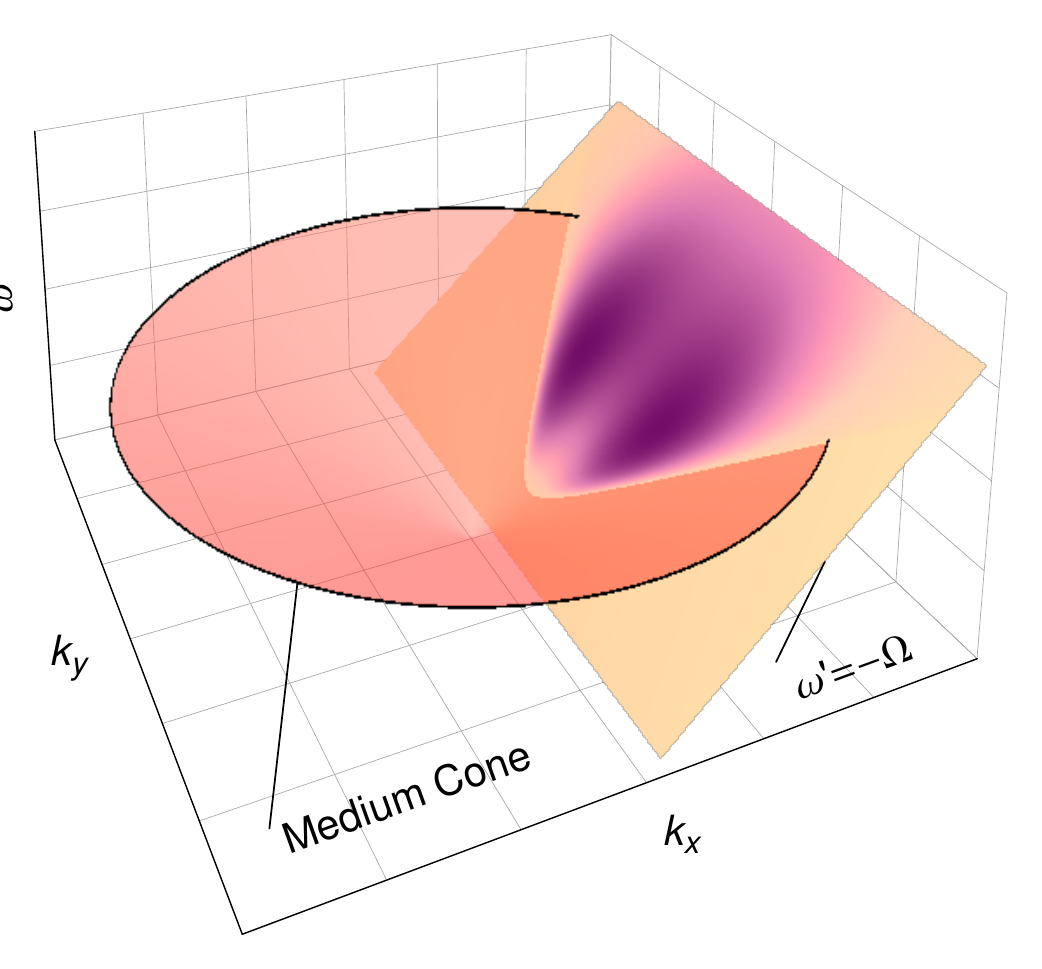}
&
	\includegraphics[width=6cm]{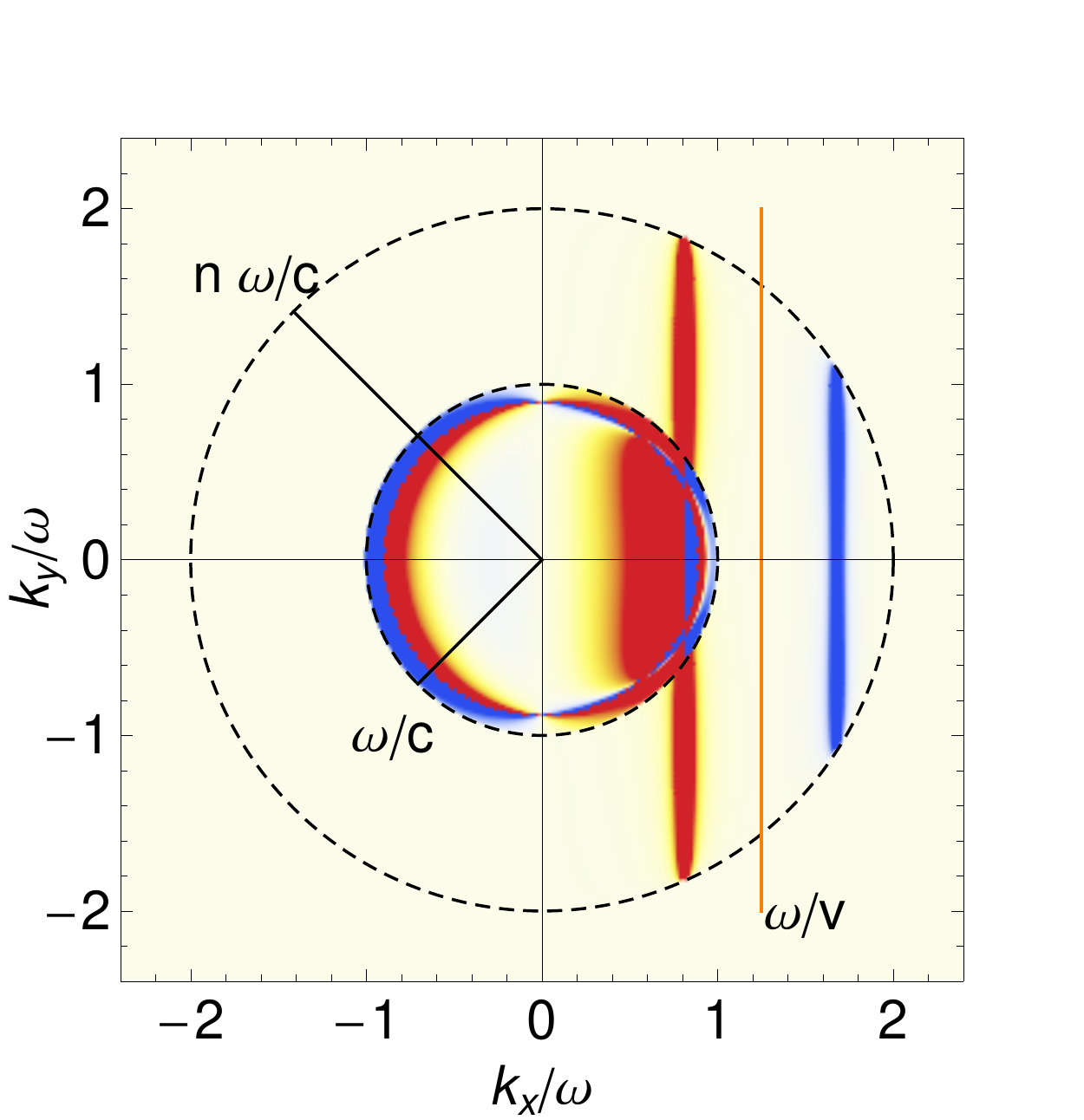}	 
\\
\multicolumn{2}{l}{%
\begin{minipage}[t]{\textwidth}
\refstepcounter{figure}%
\footnotesize{\bf Figure~\thefigure.}
Impact of the particle plasmon resonance.
\label{fig:impact-resonance}
\end{minipage}}
\\
\begin{minipage}[t]{0.5\textwidth}\footnotesize
(\emph{left})
	Spectral density $F_x(\omega,\veckparallel)$ of the friction force
	plotted in the plane $\omega' = -\Omega$ where the particle's 
	oscillator strength
	$\mathop{\rm Im}\alpha( \omega' )$ peaks. The inclination of the plane 
	is determined by the velocity of the particle. The intersection
	with the medium light cone (pink) is a hyperbola whose apex has the
	coordinates given in Eq.(\ref{eq:apex-hyperbola}).
	\label{fig:integralVisV80}
\end{minipage}
&
\begin{minipage}[t]{0.5\textwidth}\footnotesize
(\emph{right})
Density plot of $F_x(\omega,\veckparallel)$ in the $k_x,k_y$-plane
(cut through the left figure at constant frequency).
The dashed lines give the outline of the light cones in vacuum and the 
medium. The relevant integration domain is inside the larger circle
and to the right of the orange line $k_x = \omega/v$.
The wide blue line in this area illustrates the absorption resonance 
$\omega' = - \Omega$
of the particle polarizability $\alpha( \omega' )$. 
If $T \neq 0$, the orange
line blurs, and the inner circle contributes to the integral as well. Red/blue
colors represent positive/negative values.
Parameters: 
$\omega = \sqrt{3}\,\Omega$, where $\Omega$ is the particle's resonance.
We took a relatively short damping time with $\Omega \tau = 32.5$.
	\label{fig:omegaConstantSlice}
\end{minipage}
\end{tabular}
\end{figure}

Fig.\ref{fig:conesAndSlices}(\emph{right}) above
illustrates the simple appearance 
of the integration volume, which determines most of the features
of the force spectrum: it lies between
zero-frequency plane $\omega' = 0$ and the medium light cone
$\omega = \kparallel / n$. The opening angle $\phi_{\max}$ of the intersection
(measured in the $\veckparallel$-plane, relative to the direction
of the velocity ${\bf v}$) is given by the Cherenkov formula
\begin{equation}
\cos \phi_{\max} = \frac{ \omega / v }{ n \omega } = \frac{ 1 }{ n v }
\label{eq:Cherenkov-angle}
\end{equation}
where $v$ is the particle velocity (scaled to $c$).

Fig.\ref{fig:integralVisV80}(\emph{left}) shows the impact of the
particle's plasmon resonance: the plane
$\omega' = -\Omega$ and the medium light cone intersect in a
hyperbola whose opening angle (projected onto the $\veckparallel$-plane)
is again given by the Cherenkov formula~(\ref{eq:Cherenkov-angle})%
---the higher the speed of the particle, the more inclined the plane.
The integrand roughly peaks near the apex of the hyperbola
whose position is easily calculated to be
\begin{equation}
\omega_{\rm a} = 
	\frac{ \Omega }{ \gamma (n v - 1) }
~,\qquad
k_{x{\rm a}} = 
	\frac{ n \Omega / c }{ \gamma (n v - 1) }
~,\qquad
k_{y{\rm a}} = 0
~.
\label{eq:apex-hyperbola}
\end{equation}
In Fig.\ref{fig:impact-resonance}(\emph{right}), we plot a slice
at constant frequency through the  
spectral density $F_x( \omega, \veckparallel )$ of the friction force 
given by
\begin{eqnarray}
F_x(\omega,\veckparallel) &=& 
\frac{4 \hbar k_x}{\gamma (2 \pi)^3 }
	\mathop{\rm Im}\alpha[ \gamma (\omega - v k_x )]
	{\rm e}^{-2 \kappa z}
	\sum_{\sigma=s,p} 
	\frac{ 2 \kparallel q_\sigma \phi_{\sigma}(\omega, \veckparallel) 
		}{ q_\sigma^2 \kappa^2 + \kparallel^2 }
	~,
\end{eqnarray}
where the imaginary part of the reflection amplitudes
$r_\sigma$ was worked out from Eqs.(\ref{eq:reflection-coefficients}),
and $q_s = 1$ and $q_p = n^2$.
The density plot reveals how 
the resonances of the polarizability $\alpha( \omega' )$
select narrow stripes in the $\veckparallel$
plane. Only the resonance $\omega' = - \Omega$ (blue) lies in the 
integration domain relevant for quantum friction.

\begin{figure}[ht!]%
\begin{tabular}{c@{\hspace*{1em}}c}
	\includegraphics[width=7cm]{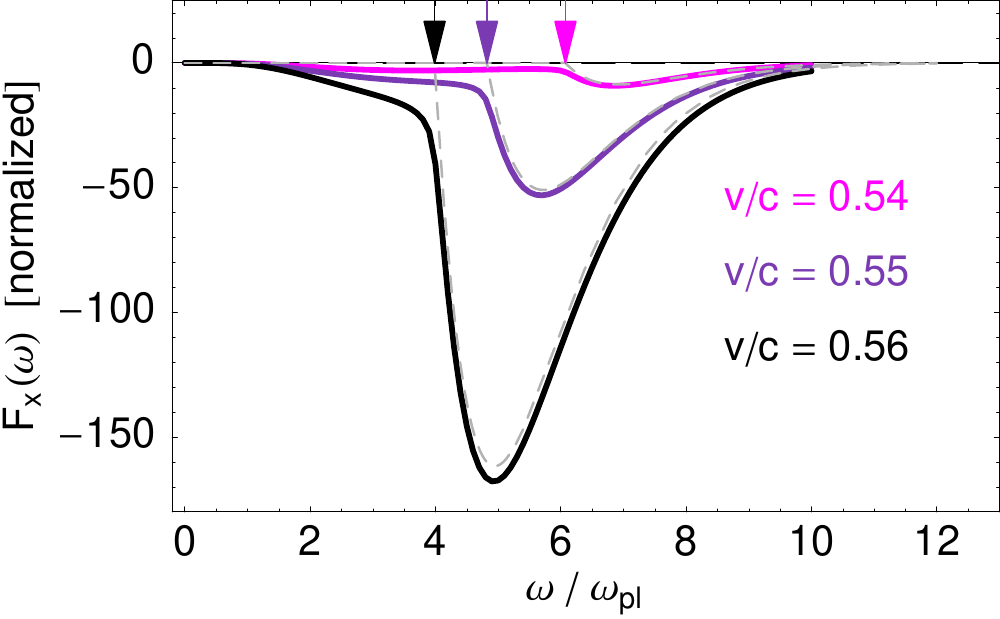}%
&
	\includegraphics[width=7cm]{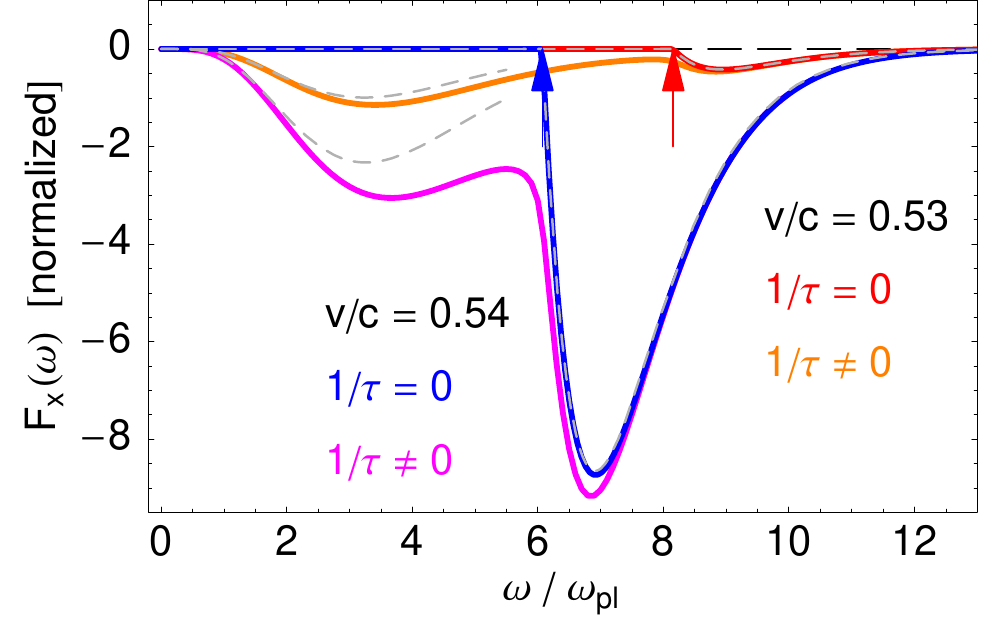}
\\
\multicolumn{2}{l}{%
\begin{minipage}[t]{\textwidth}
\refstepcounter{figure}%
\footnotesize{\bf Figure~\thefigure.}
Impact of particle velocity and plasmon damping on quantum friction.
\label{fig:accelerationSpectrum}
\label{fig:impact-of-damping}
\end{minipage}}
\\
\begin{minipage}[t]{0.5\textwidth}\footnotesize
(\emph{left})
	Frequency spectrum $F_x( \omega )$ of the friction force for a 
	silver nano-particle
	at different velocities above the Cherenkov threshold
	$c / n = 0.5$, obtained by integrating $F_x(\omega,\veckparallel)$
	over $\veckparallel$. The arrows give the apex of the 
	hyperbola~[Eq.(\ref{eq:apex-hyperbola})] shown in the
	left plot.
We used the quite arbitrary normalization factor 
$4 \hbar (4\pi a^3)(2\pi)^{-3} 10^{-4} (\omega_{\rm pl}/c)^4
\approx 3.4\,{\rm aN} / \omega_{\rm pl}$ 
for the force spectrum.
	We took a damping time fixed by $\Omega \tau = 32.5$, which is shorter than in
	a bulk due to electron scattering at the nano particle surface
	\cite{hovel1993width,kreibigvollmer1995optical,%
	berciaud2005observation,scaffardi2006sizedependence}.
\end{minipage}
&
\begin{minipage}[t]{0.5\textwidth}\footnotesize
(\emph{right})
	Comparison of the lossless case $1/\tau = 0$ and a particle 
	resonance with a finite width (same parameters as in
	Fig.\ref{fig:impact-resonance}).
	The arrow indicates the frequency $\omega_{\rm a}$
	[Eq.(\ref{eq:apex-hyperbola})]
	where the particle resonance $\omega' = - \Omega$ 
	intersects the light cone in the medium
	(apex of the hyperbola in 
	Fig.\ref{fig:integralVisV80}(\emph{left})).	
	Same normalization as in 
	Fig.\ref{fig:accelerationSpectrum}(\emph{left}).
	\label{fig:forceSpectrumNoDCompV56}
\end{minipage}
\end{tabular}
\end{figure}

These geometric considerations carry over when we 
integrate over $k_x$ and $k_y$ and consider the force spectrum. 
This is illustrated in 
Fig.\ref{fig:impact-of-damping}. Photon emission
resonant with the particle plasmon resonance becomes dominant
at velocities well above the Cherenkov threshold
[Fig.\ref{fig:impact-of-damping}(\emph{left})]. Closer to the threshold,
contributions at lower frequency arise from
photons that are off-resonant, more precisely quasi-static,
in the frame co-moving with the particle. Similar to Cherenkov
radiation, they are boosted into the visible range by the
Doppler shift. These photons arise from the nonzero value of the
polarizability at low frequencies
\begin{equation}
\omega' \ll \Omega:\quad
\mathop{\rm Im} \alpha( \omega' ) 
\approx
4\pi a^3 \frac{ \omega' }{ \Omega^2 \tau}
\label{eq:im-alpha-low-frequency}
\end{equation}
Note that the only material parameter in this regime is 
the metal conductivity $\sigma = \varepsilon_0 \Omega^2 \tau$,
see also Refs.\cite{pendry1997shearing,zhao12rotational}.
Our interpretation is confirmed in 
Fig.\ref{fig:impact-of-damping}(\emph{right})
where the spectrum is also calculated in the lossless limit,
using the approximate polarizability~(\ref{eq:imagAlphaTauInf}).
Off-resonant photon emission is suppressed, and the 
frequency $\omega_{\rm a}$ from Eq.(\ref{eq:apex-hyperbola}) 
provides a sharp threshold.

Finally, the total friction force is plotted as a function of the 
particle
velocity in Fig.\ref{fig:forceVsVelocity}.
Note again the relatively large difference between finite damping 
and the lossless limit near the Cherenkov threshold.

\subsection{Approximations near threshold}

\begin{figure}[h!]
\centerline{
	 \includegraphics[width=8cm]{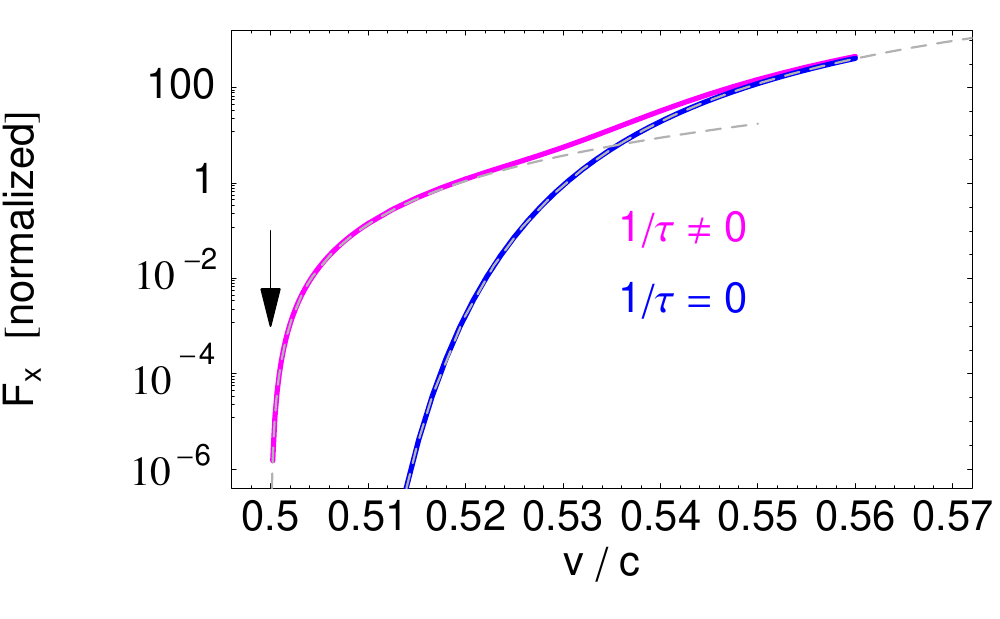}%
}
\caption[]{%
Total friction force vs.\ particle velocity (log scale). 
	The arrow indicates the Cherenkov threshold $v = c/n$.
	Similar to Fig.\ref{fig:accelerationSpectrum},
	the force is normalized to the value
	$(2/\pi^2) 10^{-4} \hbar (\omega_{\rm pl}^5/c^4) a^3$,
	i.e., an acceleration of $\approx 2900\,{\rm m/s}^2$.
}
\label{fig:forceVsVelocity}
\end{figure}

The integrals can be calculated approximately when the opening
angle of the Cherenkov cone is very narrow ($v \approx 1/n$).
The main features are captured by the reflection coefficient in
p-polarization, expanded for small $\kappa_n$ 
[see Eq.(\ref{eq:reflection-coefficients})]. 
(See the Appendix for more details.)
The formulas of this section are represented in dashed (gray) lines 
on Figs.\ref{fig:accelerationSpectrum}, \ref{fig:forceVsVelocity}: 
the agreement is quite remarkable.

For a particle polarizability with a very narrow resonance, 
we find the approximate spectrum
\begin{eqnarray}
\omega \ge \omega_{\rm a}: \quad
&&
F_x( \omega ){\rm d}\omega \approx
-\frac{ 4 \hbar  (4\pi a^3) }{ (2\pi)^3 }
\frac{ \pi^2 \Omega }{ 8 n v \gamma^2 }
{\rm d}\omega\,
\omega k_{y{\rm max}}^2 
\, {\rm e}^{ - 2 \omega z \sqrt{ n^2 - 1 } } \times
\nonumber
\\
&& \qquad {}\times
\left(4 + \frac{3 k_{y{\rm max}}^2 z }{ \omega \sqrt{n^2-1} }\right)
\label{eq:simplified-Fx-noloss-2}
	\\
&& k_{y{\rm max}}^2 =
( n v - 1 )\frac{ \omega - \omega_{\rm a} }{ v^2 }
\left[ 2 \omega + ( n v - 1 ) ( \omega + \omega_{\rm a} )
\right]
\end{eqnarray}
where $\omega_{\rm a}$ is given by Eq.(\ref{eq:apex-hyperbola}),
and $k_{y{\rm max}}$ parametrizes the width of the hyperbola
in Fig.\ref{fig:impact-resonance}(\emph{left}). This spectrum
has a sharp threshold (dashed gray lines in 
Figs.\ref{fig:accelerationSpectrum}). 
If the polarizability includes damping, the contribution
from quasi-static frequencies can be computed similarly,
using the approximation~(\ref{eq:im-alpha-low-frequency}). The
resulting spectrum is
\begin{eqnarray}
\omega \sim 0: \quad
F_x( \omega ) {\rm d}\omega &\approx&
- \frac{ 4\hbar (4\pi a^3)}{ (2\pi)^3}
\frac{ \pi  (n v-1)^3 }{ n^2 v^6 \Omega^2 \tau }
{\rm d}\omega \, \omega^5 \,{\rm e}^{-2 \omega z \sqrt{n^2-1} }
\nonumber\\
&& \qquad {} \times	
\left(
	\frac{ v^2 }{ 20 } 
	+ \frac{ 3 n v^3 }{ 20 }
	+ \frac{ 2 n^2 v^4 }{ 15 }
	\right. \label{eq:lossy-friction-spectrum}
\\
&& \qquad \left.	
   	+ \frac{ (n v - 1) \omega z }{ 280 \sqrt{n^2-1} } 
	( 5 + 20 n v + 29 n^2 v^2 + 16 n^3 v^3 )
\right)
\nonumber
\end{eqnarray}
and peaks roughly at the inverse roundtrip time
$1/(z \sqrt{n^2-1})$ (dashed lines in Fig.\ref{fig:accelerationSpectrum}(\emph{right})). As illustrated in the figure above, 
this approximation becomes quite poor away from
the threshold, as frequencies above the validity of the low-frequency 
approximation~(\ref{eq:im-alpha-low-frequency}) for
$\mathop{\rm Im}\alpha( \omega' )$ become relevant.

From both approximations for the spectra, the velocity-dependent
friction force can be calculated, leading to:
\begin{eqnarray}
\mbox{no damping}: &&
F_x \approx
-\frac{ 4\hbar (4\pi a^3) }{ (2\pi)^3 }
\frac{ \pi^2 n^3 
	}{ 8 (n^2 - 1)^{3/2} }
\frac{  \omega_{\rm a}
	}{ z^4 }
(v - 1/n)^2 {\rm e}^{-2 \sqrt{n^2-1}\, \omega_{\rm a} z } 
\nonumber\\
&& \qquad {} \times
   \left(
   3 
   + 4 \sqrt{n^2-1} \, \omega_{\rm a} z
   + 2 (n^2 - 1) (\omega_{\rm a} z)^2
   \right)
\label{eq:result-force-no-loss}
\\
\mbox{with damping}: \quad &&
F_x \approx 
- \frac{ 4\hbar (4\pi a^3) }{ (2\pi)^3 }
\frac{ 5 \pi n^5  }{ 8 ( n^2 - 1 )^3 }
\frac{ ( v - 1 / n )^3 }{ z^6 \Omega^2 \tau }
\nonumber\\
&& \qquad {} \times \left(
1 
+ ( v - 1 / n )
\frac{ 11 n - 2 n^3 }{ 4 ( n^2 - 1 ) }
+ {\cal O}(v - 1/n)^5
\right)
\label{eq:result-force-with-loss}
\end{eqnarray}
In both cases, we have simplified the complicated polynomial in
$v$ to the lowest order above $1/n$.
The dependence on the threshold frequency
$\omega_{\rm a} \sim ( v - 1/n )^{-1}$
makes the no-damping result exponentially small at threshold,
while damping leads to a cubic power law
$\sim ( v - 1/n )^{3}$.
We also emphasize the different power laws with distance $z$ from the surface;
the corrections to the no-damping case in 
Eq.(\ref{eq:result-force-no-loss})
are quite significant for our parameters, as we have the relatively 
large value $\omega_{\rm a} z \approx 2.2$ at $v = 0.55\,c$.
The numerical calculation for a particle with damping agrees quite
well with formula~(\ref{eq:result-force-with-loss}) close to the
threshold velocity. Around $v \sim 0.53$, the contribution from
the resonance takes over and the dependence on the damping constant
become negligible.

\section*{Conclusion}

We investigated a neutral particle moving in close proximity parallel to a 
dielectric. Studying the expression~(\ref{eq:generalForce}) that was derived from 
a fully relativistic extension of the fluctuation-dissipation theorem, we provided 
a connection to a fundamental and simple friction mechanism.
If the particle moves faster than the speed of light inside the medium
(Cherenkov condition),
it can dissipate energy by creating pairs of excitations.
Unlike as in Ref.\cite{barton2010van2}, the pairs are excitations 
of the particle and photon modes propagating in the medium. 
These modes change the sign of their frequency under the Doppler shift 
(anomalous Doppler
effect \cite{tamm1937coherent,ginzburg1996radiation}). This leads to
an S-matrix in the form of a Bogoliubov transformation that spontaneously
excites the particle and generates a photon emitted into the medium
\cite{frolov1986excitation,davies2005quantum,maghrebi2013bquantum}.
The mechanism we described is another example of an unstable vacuum state
in a quantum field theory \cite{silveirinha2013aquantization}.
The main features of Cherenkov friction were explained in geometrical terms 
by analyzing the frequency spectrum of the force. In order to provide a concrete example, we
considered a metallic nano particle whose polarizability is dominated
by a plasmon resonance. We found a remarkable agreement of the numerical data with an expansion of the force and the force spectrum in $(v - c/n)$ near the threshold. The approximate expressions further illustrated the role of the low frequency behavior of the particles polarizability and its plasmon resonance.

In order to connect with the current discussion on quantum 
friction \cite{philbin2009no,scheel2009,barton2010van2,intravaia2013quantum},
we note that in its simplest form, Cherenkov friction does not require 
damping in either the particle or the surface. We studied the
impact of dissipation in the particle (as described by a damped plasmon
mode) and found that this significantly changes the friction force
just above the Cherenkov velocity, while maintaining strictly zero
friction below the threshold. This changes,
however, when finite temperatures are introduced, or absorption is
allowed for in the surface. 
Our general result for the radiative force is identical to that of 
Ref.\cite{dedkov2009fluctuation}. The general setting for the field
quantization (lossless and non-dispersive dielectric) is the same
as in Ref.\cite{eberlein2006quantum}, however, a different particle
is considered there (self-energy of a moving electron).
The approach of Ref.\cite{hoye2012bcasimir} is limited to 
friction forces linear in the relative velocity of two systems which
are both at the same temperature. The vanishing of linear friction
at $T = 0$ is consistent with our analysis.
The investigation in Ref.\cite{barton2010van2} uses a different
model for the particle's polarizability: a microscopic two-level
system with radiative damping only. In the 
description of the field modes near the surface,
damping (absorption) is allowed
for, and only electrostatic fields are considered (non-relativistic
limit). We emphasize in particular that the excitations that lead 
to frictional losses are pairs of surface plasmons in 
Ref.\cite{barton2010van2}.
A comprehensive picture where
the weight of this excitation process can be compared directly
to the spontaneous particle excitation studied here still needs to be developed. 
The simple setting put forward in this paper 
may provide a route towards such a picture.

\subsection*{Acknowledgements}

We thank V. E. Mkrtchian, H. R. Haakh, and J. Schiefele for helpful
discussions in various stages of this work.

%
%
%
\section*{Appendix}
%
%
%
%
For the sake of convenience we repeat the general expression for the force at $T = 0$:
\begin{eqnarray}
F_x &=& 
	\frac{4 \hbar}{\gamma (2 \pi)^3 }
	\int\limits^\infty_0 {\rm d}\omega 
	\int\limits^{n \omega}_{\omega/v}{\rm d}k_x
	\hspace*{-1em}
	\int\limits^{\sqrt{n^2 \omega^2 - k_x^2}}_{0}
	\hspace*{-1em}
	{\rm d}k_y \nonumber
	\\
	&& \quad {} \times
	k_x\,
	{\rm Im}\alpha[ \gamma (\omega-v k_x ) ]
	\sum_{\sigma=s,p}\phi_{\sigma}(\omega,\veckparallel)
	\mathop{\rm Im}\left( r_\sigma \right) \frac{{\rm e}^{-2 \kappa z}}{\kappa}
	~,
\label{eq:}
\end{eqnarray}
Close to the threshold ($v-1/n$ becomes small) the range in $k_x$ becomes narrow ($c = 1$):
\begin{equation}
n \omega - \frac{ \omega }{ v } = 
\frac{ \omega }{ v } ( n v - 1 ) 
\approx n \omega ( n v - 1 )~.
\label{eq:range-kx}
\end{equation}
For the range in $k_y$ we find:
\begin{equation}
|k_y| \le \sqrt{ (n \omega)^2 - k_x^2 } =
\sqrt{ (n \omega - k_x)(n \omega + k_x) } \le
\omega \sqrt{ (2 n / v ) (n v - 1)  }~.
\label{eq:range-ky}
\end{equation}
Hence the wave vectors in the reflection coefficients become
\begin{eqnarray}
&& 
n^2 \omega^2 - \frac{ \omega^2 ( n^2 v^2 - 1)}{ v^2 } 
\le \kparallel^2 
\le n^2 \omega^2~,
\\
&& 
(n^2 - 1 )\omega^2 - \frac{ \omega^2 ( n^2 v^2 - 1)}{ v^2 } 
\le \kappa^2
\le (n^2 - 1 ) \omega^2~,
\\
&&
0 
\le \kappa_n^2
\le \frac{ \omega^2 ( n^2 v^2 - 1)}{ v^2 } ~.
\label{eq:range-kappa-kappa-n}
\end{eqnarray}
For small values of $\kappa_n$ the reflection coefficients thus scale like
$\mathop{\rm Im} r_\sigma \sim \sqrt{ n v - 1 }$.
For the integration domain in frequency, we distinguish 
whether the particle is lossy or lossless:
\begin{eqnarray}
\mbox{lossy:}\quad  &&
0 \le \omega < {\cal O}[ 1/(z \sqrt{n^2 - 1 }) ] < \infty~,
	\nonumber\\
\mbox{no loss:}\quad  &&
\omega_{\rm a} = \frac{ \Omega / \gamma }{ n v - 1 }
\le \omega < {\cal O}[ 1/(z \sqrt{n^2 - 1 }) ] < \infty
\label{eq:range-frequency}
\end{eqnarray}
where the upper limit arises from the exponential ${\rm e}^{ - 2 \kappa z }$.
For the comoving frequency  $\omega' = \gamma( \omega - v k_x )$ we find:
\begin{eqnarray}
\mbox{lossy:}\quad  &&
0 \le - \omega' \le
\gamma \omega (n v - 1 ) < 
{\cal O}[ \gamma (n v - 1 )/(z \sqrt{n^2 - 1 }) ]~,
	\nonumber\\
\mbox{no loss:}\quad  &&
- \omega' = \Omega~.
\label{eq:}
\end{eqnarray}
Hence we see how the relevant frequency ranges are separated by a considerable margin between the lossy and the lossless case. This
suggests that we can capture both from two different approximations.
Note that the lossy case is dominated by a range of
quasi-static frequencies which becomes narrower, the closer the
velocity gets to the Cherenkov threshold (difference $v - 1/n$).
%
%
%
\section*{Particle with no losses}
%
%
%
In the case of no losses we approximate the particle's resonance with a $\delta$-distribution.
Formally this is done by taking the limit $1/\tau\rightarrow 0$
in Eq.(\ref{eq:particle-polarizability}).
Because we only have to focus on $\omega < 0$ we find
[see also Eq.(\ref{eq:imagAlphaTauInf})]:
\begin{eqnarray}
&& 
(4\pi a^3)
\mathop{\rm Im} 
\frac{ \Omega^2 }{ \Omega^2 - \omega^2 - {\rm i} \omega / \tau }
\approx
- (4\pi a^3) \frac{ \pi }{ 2 } \delta( \omega + \Omega )~.
\label{eq:check-delta}
\end{eqnarray}
The oscillator strength $\mathop{\rm Im}\alpha( \omega' )$ thus
fixes the $k_x$-wave vector to (label `a' from `apex' of hyperbola)
\begin{equation}
\mbox{no loss:}\quad
k_x = k_{x{\rm a}} = \frac{ \omega + \Omega/\gamma }{ v }
\,, \qquad
\omega > \omega_{\rm a} = \frac{ \Omega / \gamma }{ n v - 1 }~.
\label{eq:def-kxa}
\end{equation}
Note that $\omega$ becomes `large' near the threshold $n v = 1$.
The expressions for wave vectors $\kappa$ and $\kappa_{n}$ read
\begin{eqnarray}
k_{y{\rm max}}^2 &=& (n\omega)^2 - k_{x{\rm a}}^2 ~,
\\
\kappa_{n}^2 & = &
k_{y{\rm max}}^2 - k_y^2 ~,
\\
\kappa & \approx & 
\sqrt{n^2 - 1}\,\omega - 
\frac{ k_{y{\rm max}}^2 - k_y^2 }{ 2 \sqrt{n^2 - 1}\,\omega }~.
\label{eq:}
\end{eqnarray}
For the reflection coefficients this yields the approximation
(keeping only the lowest order in $\kappa$)
\begin{equation}
\mathop{\rm Im} r_p \approx
\frac{ 2 }{ n^2 \sqrt{ n^2 - 1 } }
\frac{ k_{y{\rm max}} }{ \omega }  (1 - q_y^2)^{1/2}
\label{eq:}
\end{equation}
where $0 \le q_y \le 1$ is a scaled version of $k_y = 
q_y k_{y{\rm max}}$. For the exponential, we include the next order 
of $\kappa$ and take
\begin{equation}
\frac{ {\rm e}^{ - 2 \kappa z } }{ \kappa }
\approx
\frac{ {\rm e}^{ - 2 \omega z \sqrt{ n^2 - 1 } } }{ \omega \sqrt{ n^2 - 1 } }
\left[ 1 +
\frac{ z k_{y{\rm max}}^2 }{ \sqrt{n^2 - 1}\,\omega }
( 1 - q_y^2)
\right]~.
\label{eq:}
\end{equation}
We find that the p-polarization yields the leading contribution
and use the lowest-order approximation
\begin{eqnarray}
\phi_p( \omega, \veckparallel ) & = &
2 \omega^2 ( n^2 - 1 ) + {\cal O}[ \omega^2( n v - 1 ) ]~,
\\
\phi_s( \omega, \veckparallel ) & = &
{\cal O}[ \omega^2 (n v - 1 ) ]~.
\label{eq:}
\end{eqnarray}
Putting everything together in the leading order, we get the approximate
expression 
\begin{eqnarray}
F_x &\approx&
-\frac{ 4 \hbar (4\pi a^3)}{ (2\pi)^3 }
\frac{ 2 \pi \Omega }{ n v \gamma^2 }
\int\limits_{\omega_{\rm a}}^\infty\!{\rm d}\omega\,
\omega k_{y{\rm max}}^2 
\, {\rm e}^{ - 2 \omega z \sqrt{ n^2 - 1 } }
\nonumber
\\
&& \qquad {}\times
\int\limits^{1}_{0}\!{\rm d}q_y\,
(1 - q_y^2)^{1/2}
\left[ 1 + 
\frac{ z k_{y{\rm max}}^2 }{ \omega \sqrt{n^2 - 1} }
( 1 - q_y^2)
\right]~.
\label{eq:}
\end{eqnarray}
A convenient formulation for
\begin{equation}
k_{y{\rm max}}^2 = 
( n v - 1 )\frac{ \omega - \omega_{\rm a} }{ v^2 }
\left[ 2 \omega + ( n v - 1 ) ( \omega + \omega_{\rm a} )
\right]
\label{eq:}
\end{equation}
shows the scaling above threshold.
The $q_y$-integral can be performed and produces Eq.(\ref{eq:simplified-Fx-noloss-2}).
%
%
\section*{Particle with losses}
%
%
As outlined in the estimates above, we can use the low-frequency 
approximation
\begin{equation}
\omega' \ll \Omega:\quad
\mathop{\rm Im} \alpha( \omega' ) 
\approx
4\pi a^3 \frac{ \omega' }{ \Omega^2 \tau}~.
\label{eq:}
\end{equation}
for the polarizability. This assumes that the frequency range
for $\omega'$ near zero is sufficient to capture the integral and
ignores the resonant peak. The integrals that must be performed are
\begin{eqnarray}
F_x &\approx& - \frac{ 4\hbar (4\pi a^3)}{ (2\pi)^3}
\frac{ 4 n^2 }{ \Omega^2 \tau } 
\int\limits_0^\infty\!{\rm d}\omega \,
\omega^5 
\,{\rm e}^{ - 2 \omega z \sqrt{ n^2 - 1 } }
\nonumber\\
&& \qquad
\int\limits_{1/(nv)}^1\!{\rm d}q_x \,
q_x 
( 1 - q_x^2 )
( v n q_x - 1 )
 \nonumber\\
&& \qquad
\int\limits_{0}^1\!{\rm d}q_y \,
\sqrt{ 1 - q_y^2 } 
\left[ 1 + 
\frac{ \omega z n^2 ( 1 - q_x^2 ) }{ \sqrt{n^2 - 1} }
( 1 - q_y^2)
\right] ~.
\label{eq:}
\end{eqnarray}
The $q_y$-integral is the same as before [Eq.(\ref{eq:simplified-Fx-noloss-2})] 
and gives
\begin{eqnarray}
F_x &\approx& - \frac{ 4\hbar (4\pi a^3)}{ (2\pi)^3}
\frac{ 4 n^2 }{ \Omega^2 \tau } 
\int\limits_0^\infty\!{\rm d}\omega \,
\omega^5
\,{\rm e}^{ - 2 \omega z \sqrt{ n^2 - 1 } }
\nonumber\\
&& \qquad
\int\limits_{1/(nv)}^1\!{\rm d}q_x \,
q_x 
( 1 - q_x^2 )
( v n q_x - 1 )
\frac{ \pi }{ 4 }
\left( 1 + 
\frac{ 3 \omega z n^2 ( 1 - q_x^2 ) }{ 4 \sqrt{n^2 - 1} }
\right) ~.
\label{eq:}
\end{eqnarray}
The $q_x$-integral yields the force spectrum in eq.(\ref{eq:lossy-friction-spectrum}).

%
\newpage

\bibliographystyle{unsrt}
\bibliography{mylib}

\end{document}